\documentclass{ws-procs961x669}            
\begin{document}
\title{$U(1)$ local strings in hybrid metric-Palatini gravity}

\author{Tiberiu Harko}

\address{Astronomical Observatory, 19 Ciresilor Street, 400487 Cluj-Napoca, Romania \\
Faculty of Physics, Babes-Bolyai University, 1 Kogalniceanu Street, 400084 Cluj-Napoca, Romania \\
School of Physics, Sun Yat-Sen University, Xingang Road,
510275 Guangzhou, People’s Republic of China}

\author{Francisco S. N. Lobo}
\address{Instituto de Astrofísica e Ciências do Espaço, Faculdade de Ciências da Universidade de Lisboa,
Edifício C8, Campo Grande, P-1749-016 Lisbon, Portugal}

\author{Hilberto M. R. da Silva}

\address{Instituto de Astrofísica e Ciências do Espaço, Universidade do Porto, CAUP, Rua das Estrelas,
PT4150-762 Porto, Portugal and Centro de Astrofísica da Universidade do Porto,
Rua das Estrelas, PT4150-762 Porto, Portugal}

\begin{abstract}
In this work we made use of a general static cillindrically symmetric metric to find $U(1)$ local cosmic string solutions in the context of the hybrid metric-Palatini theory of gravity in it's scalar-tensor representation. After finding the dynamical equations for this particular case, we imposed boost invariance along $t$ and $z$ directions, which simplified the equations of motions, leaving only one single metric tensor component, $W^2(r)$. For an arbitrary potential $V(\phi)$, the solutions obtained can be put in a closed parametric form, with $\phi$ taken as a parameter. Several particular cases of the potential were studied, some yielding simple mathematical forms, others with only numerical solutions. With this approach, we obtain a large number of new stable stringlike solutions in hybrid metric-Palatini gravity, in which the parameters, like the scalar field, metric tensor components, and string tension, depend fundamentally on the boundary values of the scalar field, and its derivative, on the $r(0)$ circular axis.
\end{abstract}

\keywords{Cosmic Strings, Modified Gravity, hybrid metric-Palatini theory}

\bodymatter

\section{Introduction}\label{intro}

The recent observation of the accelerated expansion of the Universe \cite{SupernovaSearchTeam:1998fmf} brought an identity crisis upon the standard theory of gravity, General Relativity. Albeit being a well established theory, with several important experimental confirmations, there is now a wide agreement that General Relativity may be, at the very least, an incomplete theory \cite{Clifton:2011jh}. Not only due to the observations that demand the introduction of a Dark Sector to the energy budget of the Universe, but also due to theoretical difficulties, namely the existence of mathematical singularities and the non-quantization of the theory \cite{Utiyama:1962sn}. All of these issues paved the way to more general theories for gravitation, focusing on solving some of these problems, or at least to be used to falsify the theory. Lovelock's theorem \cite{Lovelock:1971yv} is a crucial roadmap to this quest, as it establishes the GR equations as the only possible solution (minus constant factors), if one assumes the same conditions Einstein followed when he proposed the theory. In order to produce a viable modified gravity theory, at least one of the assumptions must be broken.

The hybrid metric-Palatini theory was proposed by Capozzielo, Harko, Koivisto, Lobo and Olmo in 2011 \cite{Harko:2011nh} in order to overcome the difficulties faced by $f(R)$ theories of gravity and successfully unifies the late-time cosmic acceleration period with the weak-field solar system dynamics without the need for chameleon mechanisms \cite{Harko:2011nh}. For a more in-depth review of the theory, we refer the reader to the review article \cite{Harko:2020ibn}. In this theory, the connection is considered an independent dynamical variable, adding to the metric.

Cosmic strings are one of the possible topological defects formed after spontaneous symmetry breaking (SSB) during phase transitions in the history of the Universe. In fact, many grand
unification theories (GUTs) postulate that our universe, as it
cooled, underwent a series of phase transitions related
with SSB, meaning that at a higher temperature
there was invariance under a more general group of
symmetries. Each of these phase transitions may be responsible for the formation of a network of topological defects, through the Kibble-Zurek mechanism. For a more in depth review of topological defects, we refer the reader to Ref.\citenum{Vilenkin:2000jqa}

The type of strings to be considered in this work are local $U(1)$ cosmic strings, which are an extension of the global $U(1)$ strings to include gauge fields. Local strings differ from the global cosmic strings in what concerns the symmetry that is effective above the spontaneous breaking scale; in the case of local strings, the lagrangian remains invariant under local transformations of the type $\phi(x) \longrightarrow e^{i\alpha(x)}\phi (x)$.

The study of the properties and dynamics of cosmic strings in the context of modified theories of gravity is crucial in the advent of powerful observatories, such as LISA, as it may allow us to constrain both Modified Gravity theories and Grand Unified theories.

\section{Hybrid metric-Palatini Gravity}
In this hybrid regime, the metric and Palatini approaches are combined, by adding a new term, $f(\mathcal{R})$ to the Einstein-Hilbert action:
\begin{equation}\label{actionhmp}
    S=\frac{1}{2\kappa^2}\int\sqrt{-g}\ d^4x\ \left[R+f(\mathcal{R}) +\mathcal{L}_m\right].
\end{equation}
where $\mathcal{R}$ is the Ricci-Palatini scalar, the geometrical analogous to the Ricci scalar, constructed in terms of an independent connection, $\hat{\Gamma}$:
\begin{equation}\label{palatiniriccitensor}
    \mathcal R_{\mu\nu}=\partial_\alpha\hat{\Gamma}^\alpha_{\mu\nu}-\partial_\nu\hat{\Gamma}^\alpha_{\mu \alpha}+\hat{\Gamma}^\alpha_{\alpha \beta}\hat{\Gamma}^\beta_{\mu \nu}-\hat{\Gamma}^\alpha_{\mu \beta}\hat{\Gamma}^\beta_{\alpha \nu}.
\end{equation}
\begin{equation}\label{palatiniricciscalar1}
    \mathcal{R}=g^{\mu\nu}\mathcal{R}_{\mu\nu}
\end{equation}
In this theory we have two dynamical variables, the metric and the (independent) connection. And so, we apply variation calculations of the action with respect to both.

Varying \eqref{actionhmp} with respect to the metric, $g_{\mu\nu}$ we obtain
\begin{equation}\label{dynamichmp}
    G_{\mu \nu} + \frac{df}{d\mathcal{R}} \mathcal{R}_{\mu\nu} - \frac{1}{2}f(\mathcal{R})g_{\mu\nu}=\kappa^2T_{\mu\nu}
\end{equation}
where $T_{\mu\nu}$ is the energy-momentum tensor defined, as usual, as:
\begin{equation}\label{energytensor}
    T_{\mu\nu} \equiv -\frac{2}{\sqrt{-g}}\frac{\delta(\sqrt{-g}\mathcal{L}_m)}{\delta(g^{\mu\nu})}
\end{equation}

When we perform the variation of the action \eqref{actionhmp} with respect to the independent connection, $\hat{\Gamma}$, the resulting equation is:
\begin{equation}\label{dynamic2hmp}
    \hat{\nabla}_\alpha=\left( \sqrt{-g}\frac{df}{d\mathcal{R}}g^{\mu\nu}\right)
\end{equation}
which has an interesting feature: if we consider a metric conformally related to $g_{\mu\nu}$ by a $\frac{df}{d\mathcal{R}}$ factor, $h_{\mu\nu}=\frac{df}{d\mathcal{R}}g_{\mu\nu}$, equation \eqref{dynamic2hmp} implies that the independent connection $\hat{\Gamma}$ is the Levi-Civita connection of such metric.

One of the interesting feature of the hybrid metric-Palatini theory is that it admits a scalar-tensor representation, which simplifies the analysis of the dynamics of the theory.

Let's introduce an auxiliary field, $A$, so that the action \eqref{actionhmp} becomes
\begin{equation}
\label{actionhmp2}
    S=\frac{1}{2\kappa^2}\int\sqrt{-g}\ d^4x \left[\Omega_A R+f(A) + \frac{df}{dA}(\mathcal{R}-A) +\mathcal{L}_m\right].
\end{equation}
where the coupling constant $\Omega_A$ is introduced for generality. If we further define $\phi \equiv \frac{df}{dA}$ and $V(\phi)\equiv A \frac{df}{dA}-f(A)$, the action becomes:
\begin{equation}\label{actionhmp3}
    S=\frac{1}{2\kappa^2}\int\sqrt{-g}\ d^4x\left[\Omega_A R+\phi \mathcal{R} - V(\phi) +\mathcal{L}_m\right]\ .
\end{equation}

Since we now have three dynamical variables, the metric $g_{\mu\nu}$, the independent connection, $\hat{\Gamma}$ and the scalar field $\phi$, we perform the variation of the action \eqref{actionhmp3} with respect to each of the dynamical variables, resulting in the following equations of motion:
\begin{equation}\label{motion1}
    \Omega_A R_{\mu\nu} + \phi \mathcal{R}_{\mu\nu} -\frac{1}{2}(\Omega_A R + \phi \mathcal{R} - V(\phi))g_{\mu\nu} = \kappa^2 T_{\mu\nu}
\end{equation}
\begin{equation}\label{motion2}
    \mathcal{R}-V_\phi=0
\end{equation}
\begin{equation}\label{motion3}
    \hat{\nabla}_\alpha=\left( \sqrt{-g}\phi g_{\mu\nu}\right)
\end{equation}

As we've seen before, the last equation implies that the independent connection is the Levi-Civita connection to a conformal related metric to $g_{\mu\nu}$: $h_{\mu\nu}=\phi g_{\mu\nu}$, allowing us to write the Ricci-Palatini tensor from the Ricci tensor and the conformal factor $\phi$:
\begin{equation}\label{palatiniriccitensor2}
    \mathcal{R}_{\mu\nu}=R_{\mu\nu} - \frac{1}{\phi} \left( \nabla_\mu \nabla_\nu \phi + \frac{1}{2}g_{\mu\nu}\Box \phi \right) + \frac{3}{2\phi^2}\partial_\mu \phi \partial_\nu \phi
\end{equation}
\begin{equation}
    \label{palatiniricciscalar}
    R=\mathcal{R} + \frac{3}{\phi}\Box \phi - \frac{3}{2\phi^2}\partial_\mu \phi \partial^\mu \phi 
\end{equation}
where $\Box \equiv \nabla_\mu\nabla^\mu$ in the D'Alembert operator. Eq.\eqref{palatiniricciscalar} can be used to recast the action \eqref{actionhmp3} into the following form:
\begin{equation}\label{actionhmp4}
    S=\frac{1}{2\kappa^2}\int\sqrt{-g}\ d^4x\left[(\Omega_A +\phi)R +  \frac{3}{2\phi}\partial_\mu \phi \partial^\mu \phi - V(\phi) +\mathcal{L}_m\right]\ .
\end{equation}

We can note now three different cases for the value of the coupling constant $\Omega_A$, for $\Omega_A=1$ we have the original hybrid threory, for $\Omega_A=0$ we have the Palatini-$f(\mathcal{R})$ gravity and for $\Omega_A\rightarrow \infty$ we recover the metric-$f(R)$ theory.

Using eqs.\eqref{palatiniriccitensor} and \eqref{motion2}, we can rewrite eq.\eqref{motion1} in the scalar-tensor representation for the original hybrid theory as:
\begin{equation}\label{fieldequationhmp}
    (1+\phi)R_{\mu\nu}=\kappa^2 \left( T_{\mu\nu}-\frac{1}{2}g_{\mu\nu}T \right) + \frac{1}{2}g_{\mu\nu}(V+\Box \phi)+ \nabla_\mu \nabla_\nu \phi - \frac{3}{2\phi}\partial_\mu \phi \partial_\nu \phi
\end{equation}

In this equation we can conclude that the curvature of spacetime is due not only to the matter distribution, but also to the presence of the scalar field.

Taking the trace of equation \eqref{motion1} with $g^{\mu\nu}$, and using equation \eqref{motion2}, we get
\begin{equation}
\label{thisequationhere}
    2V-\phi V_\phi=\kappa^2 T + \Omega_A R
\end{equation}

If we now rewrite eq. \eqref{thisequationhere} using eqs. \eqref{palatiniricciscalar} and \eqref{motion2}, we get an effective Klein-Gordon equation for the scalar field:
\begin{equation}\label{KG}
-\Box \phi +\frac{1}{2\phi}\partial_\mu \phi \partial^\mu\phi+\frac{\phi\left[2V-(1+\phi)V_\phi\right]}{3}=\frac{\phi \kappa^2}{3}T
\end{equation}

This equation shows that, unlike the purely Palatini case, the scalar field is dynamic and the theory is therefore not affected by the instabilities found in the Palatini gravity \cite{Olmo:2011uz}.

\section{Dynamical equations of local strings in Hybrid metric-palatini using Vilenkin's approximation}

For the study of local $U(1)$ string in the context of the hybrid metric-Palatini gravity, we will start by defining the energy-momentum tensor of a straight, infinite cosmic string along the $z$-direction. Following Vilenkin's prescription \cite{Vilenkin:1981zs}, one possible energy-momentum tensor is 
\begin{equation}\label{energymomentumtensorstring}
    T_r^r=T_t^t=-\sigma(r)
\end{equation}
where $\sigma(r)$ is the linear energy density of the string.

We consider a general cylindrically symmetric static metric
\begin{equation}\label{metric}
    ds^2=-e^{2(K-U)}dt^2+e^{2(K-U)}dr^2+e^{-2U}W^2d\theta^2+e^{2U}dz^2
\end{equation}
Where $t$, $r$, $\theta$ and $z$ denote the time, radial, angular and axial cylindrical coordinates respectively, and $K$, $U$ and $W$ are functions of $r$ alone.

Inserting the metric \eqref{metric} into the gravitational field equation \eqref{fieldequationhmp} provides the following components
\begin{eqnarray}\label{tt}
    (1+\phi)\left(-U'^2+K' \frac{W'}{W}-\frac{W''}{W}\right)= \phi'' - \frac{3}{4\phi}\phi'^2
	\nonumber  \\    
    - \left( K'-U' - \frac{W'}{W}  \right)\phi'   + \left( \kappa^2\sigma + \frac{1}{2}V \right) e^{2(K-U)} \,,
\end{eqnarray}

%
\begin{equation}\label{rr}
    (1+\phi) \left(-U'^2+K' \frac{W'}{W} \right)=- \frac{3}{4\phi}\phi'^2
    - \left( K'-U' +\frac{W'}{W} \right)\phi'   -  \frac{1}{2}V  e^{2(K-U)} \,,
\end{equation}
%

\begin{equation}\label{thetatheta}
    (1+\phi) \left(U'^2+K'' \right)=-\phi''+\frac{3}{4\phi}\phi'^2 -U'\phi'    -\frac{1}{2}Ve^{2(K-U)}  \,,
\end{equation}
%

\begin{eqnarray}\label{zz}
&&    (1+\phi) \left(U'^2 + K'' -2U'' -2U'\frac{W'}{W}+\frac{W''}{W} \right)= -\phi'' 
	\nonumber  \\    
&&    + \frac{3}{4\phi}\phi'^2
    + \left( U' - \frac{W'}{W}  \right)\phi'   -  \left( \kappa^2 \sigma + \frac{1}{2}V \right) e^{2(K-U)}.
\end{eqnarray}

Additionally, we can use Eq. \eqref{KG} to determine the effective Klein-Gordon equation for the scalar field $\phi$:
\begin{equation}\label{KG1}
    e^{-2(K-U)} \left(-\phi''-\frac{W'}{W}\phi'+\frac{\phi'^2}{2\phi} \right)+\frac{\phi}{3}\left[2V-(\phi+1)V_{,\phi}\right]\frac{2\phi\kappa^2\sigma}{3}=0 \,.
\end{equation}

As in this model the matter field and curvature couple only minimally, it's possible to show that the energy conservation equation still holds, i.e.,
\begin{equation}\label{conservation}
    \nabla_\mu T^{\mu}{}_{\nu}=0
\end{equation}
which yields $K'\sigma=0$, this implies$K'=0$ (or the trivial vacuum $\sigma=0$). Thus,  we consider from now on that $e^K=1$, so that Eqs. \eqref{tt}--\eqref{zz} simplify to the following relations
\begin{eqnarray}\label{tt1}
    (1+\phi)\left(-U'^2-\frac{W''}{W}\right)&=& \phi'' - \frac{3}{4\phi}\phi'^2
    + \left( U' + \frac{W'}{W}  \right)\phi'   
	\nonumber  \\    
 &+& \left( \kappa^2\sigma + \frac{1}{2}V \right) e^{-2U} \,,
\end{eqnarray}

\begin{equation}\label{rr1}
    (1+\phi) U'^2 = \frac{3}{4\phi}\phi'^2
     +\left( -U' +\frac{W'}{W} \right)\phi'  +   \frac{1}{2}V  e^{-2U} \,,
\end{equation}

\begin{equation}\label{thetatheta1}
    (1+\phi) U'^2 =-\phi''+\frac{3}{4\phi}\phi'^2 -U'\phi'    -\frac{1}{2}Ve^{-2U} \,,
\end{equation}

\begin{eqnarray}\label{zz1}
&&    (1+\phi) \left(U'^2 -2U'' -2U'\frac{W'}{W}+\frac{W''}{W} \right)= -\phi'' + \frac{3}{4\phi}\phi'^2
	\nonumber \\    
&&  \qquad \qquad   + \left( U' - \frac{W'}{W}  \right)\phi'   -  \left( \kappa^2 \sigma + \frac{1}{2}V \right) e^{-2U} \,,
\end{eqnarray}
With the effective Klein-Gordon equation for the scalar field $\phi$ reducing to
\begin{eqnarray}\label{KG2}
    e^{2U} \left(-\phi''-\frac{W'}{W}\phi'+\frac{\phi'^2}{2\phi} \right)+\frac{\phi}{3}\left[2V-(\phi+1)V_{,\phi}\right]+\frac{2\phi\kappa^2\sigma}{3}=0 \,.
\end{eqnarray}

If we further consider that local gauge strings preserve boost invariance along the $t$ and $z$ directions  \cite{Vilenkin:1981zs}, meaning, in this case, that $U=0$, the gravitational field equations simplify considerably:
\begin{equation}\label{tt2}
(1+\phi)\left(-\frac{W''}{W}\right)= \phi'' - \frac{3}{4\phi}\phi'^2
+\frac{W'}{W}  \phi'   + \kappa^2\sigma + \frac{1}{2}V \,,
\end{equation}
\begin{equation}\label{rr2}
0= \frac{3}{4\phi}\phi'^2 +\frac{W'}{W}\phi'  +   \frac{1}{2}V  \,,
\end{equation}
\begin{equation}\label{thetatheta2}
0=-\phi''+\frac{3}{4\phi}\phi'^2     -\frac{1}{2}V  \,,
\end{equation}
\begin{equation}\label{zz2}
(1+\phi) \frac{W''}{W} = -\phi'' + \frac{3}{4\phi}\phi'^2
 - \frac{W'}{W}  \phi'   -   \kappa^2 \sigma - \frac{1}{2}V \,.
\end{equation}
where we can see now that Eqs. \eqref{tt2} and \eqref{zz2} become redundant.
Combining Eqs. \eqref{rr2} and \eqref{thetatheta2} yields the following relation for the potential $V$:
\begin{equation}\label{simp2}
V=-\phi''-\frac{W'}{W}\phi'  \,,
\end{equation}
substituting into the Klein-Gordon equation \eqref{KG2}, the latter reduces to:
\begin{equation}\label{KGsimp1}
V\left( 3+2\phi\right) -V_{\phi}\phi\left( \phi+1\right) +2\kappa^2\sigma\phi + \frac{3\phi'^2}{2\phi}=0 \,.
\end{equation}
Additionally, we can further deduce:
\begin{equation}\label{sigma1}
\kappa^2\sigma=\frac{1}{W}\left[(1+\phi)W'\right]' \,,
\end{equation}
and
\begin{equation}\label{this}
\frac{\left[(1+\phi)W\right]''}{W}=-(V+\kappa^2\sigma) \,.
\end{equation}

An important physical parameter characterizing the cosmic string properties is the mass per unit length of the string, which is defined as
\begin{equation}
\begin{split}
m(r)&=\int_0^{2\pi}{d\theta }\int_0^{R_s}{\sigma (r)W(r)dr} \\
&=2\pi \int_0^{R_s}{\sigma (r)W(r)dr},
\end{split}
\end{equation}
where $R_s$ is the string radius.

This set of equations allow us to write the gravitational equations of a cosmic string in hybrid metric-Palatini gravity in an exact(closed) form, where all the geometric and physical quantities are given in a parametric form, with $\phi$ taken as the parameter.

By taking into account Eq.~\eqref{thetatheta2}, the field equations \eqref{tt2} and \eqref{zz2} reduce to the form
\begin{equation}\label{de0}
\left(1+\phi\right)\frac{W''}{W}=-\frac{W'}{W}\phi '-\kappa ^2\sigma.
\end{equation}
\subsection{Parametric form of the dynamical equations}

 From a mathematical point of view Equation \eqref{thetatheta2} is independent of the metric tensor coefficient $W$ and it represents a second order nonlinear differential equation. In order to solve it we first rescale the radial coordinate $r$ according to the transformation $r=\beta \xi$. Hence Eq.~\eqref{thetatheta2} takes the form
\begin{equation}
\label{de1}
\frac{d^2\phi}{d\xi^2}-\frac{3}{4\phi}\left(\frac{d\phi}{d\xi}\right)^2+\frac{1}{2}\beta ^2V(\phi)=0.
\end{equation}

In order to solve Eq.~(\ref{de1}) we introduce the transformations
\begin{equation}
\frac{d\phi}{d\xi}=u, \quad \frac{d^2\phi}{d\xi ^2}=\frac{du}{d\xi}=\frac{du}{d\phi}\frac{d\phi}{d\xi}=u\frac{du}{d\phi}=\frac{1}{2}\frac{d}{d\phi}u^2,
\end{equation}
and
\begin{equation}
u^2=v,
\end{equation}
respectively. Then Eq.~(\ref{de1}) becomes a first order linear differential equation of the form
\begin{equation}
\frac{dv}{d\phi}-\frac{3}{2\phi}v+\beta ^2V(\phi)=0,
\end{equation}
with the general solution given by
\begin{equation}
v(\phi)=\phi ^{3/2}\left[C-\beta ^2\int{\phi ^{-3/2}V(\phi)d\phi}\right],
\end{equation}
where $C$ is an arbitrary constant of integration. We immediately obtain
\begin{equation}
u(\phi)=\phi ^{3/4}\sqrt{\left[C-\beta ^2\int{\phi ^{-3/2}V(\phi)d\phi}\right]},
\end{equation}
and
\begin{equation}\label{xi}
\xi +C _0=\int{\frac{\phi ^{-3/4}d\phi}{\sqrt{\left[C-\beta ^2\int{\phi ^{-3/2}V(\phi)d\phi}\right]}}},
\end{equation}
respectively, where $C _0$ is an arbitrary constant of integration. 

Equation~(\ref{rr2}) can be successively transformed as
\begin{equation}
\frac{1}{W}\frac{dW}{d\xi}\frac{d\phi}{d\xi}=-\frac{3}{4\phi}\left(\frac{d\phi}{d\xi}\right)^2-\frac{\beta ^2}{2}V(\phi),
\end{equation}
and
\begin{equation}
\begin{split}
\frac{1}{W}\frac{dW}{d\phi}=&-\frac{3}{4\phi}-\frac{\beta^2}{2}\frac{\phi ^{-3/2}V(\phi)}{\left[C-\beta^2\int{\phi^{-3/2}V(\phi)d\phi}\right]}\qquad \\=& -\frac{3}{4\phi}+\frac{1}{2}\frac{d}{d\phi}\ln\left[C-\beta^2\int{\phi ^{-3/2}V(\phi)d\phi}\right],
\end{split}
\end{equation}
yielding
\begin{equation}\label{W}
W(\phi)=W_0\phi ^{-3/4}\sqrt{C-\beta ^2\int{\phi ^{-3/2}V(\phi)d\phi}},
\end{equation}
where $W_0$ is an arbitrary constant of integration. 

As a last step we need to obtain the expression of $\sigma$. With the use of Eq.~(\ref{rr2}), then Eq.~(\ref{de0}) can be rewritten as
\begin{equation}\label{47}
\left(1+\phi\right)\frac{1}{W}\frac{d^2W}{d\xi^2}=\frac{3}{4\phi}\left(\frac{d\phi}{d\xi}\right)^2+\frac{1}{2}\beta ^2V(\phi)-\beta ^2\kappa ^2\sigma.
\end{equation}
With the use of the mathematical identities
\begin{equation}
\frac{dW}{d\xi}=\frac{dW}{d\phi}\frac{d\phi}{d\xi}=\frac{dW}{d\phi}u,
\end{equation}
\begin{equation}
\frac{d^2W}{d\xi ^2}=\frac{d^2W}{d\phi ^2}v+\frac{1}{2}\frac{dW}{d\phi}\frac{dv}{d\phi},
\end{equation}
Eq. (\ref{47}) takes the form
\begin{equation}
\begin{split}
(1+\phi)\left(\frac{1}{W}\frac{d^2W}{d\phi ^2}v+\frac{1}{2}\frac{1}{W}\frac{dW}{d\phi}\frac{dv}{d\phi}\right)
=\frac{3}{4\phi}v+\frac{1}{2}\beta ^2V(\phi)-\beta ^2\kappa ^2\sigma.
\end{split}
\end{equation}
Finally, after some simple calculations we obtain
\begin{equation}\label{sigma}
\begin{split}
\kappa ^2\sigma (\phi)=& \frac{1}{4 \phi} \left\{ \Big[2 \phi  (\phi +1) V'(\phi )+3 \sqrt{\phi } \int \frac{V(\phi )}{\phi^{3/2}} \, d\phi \right.\\	
&\left. -2 (2 \phi +3) V(\phi )\Big]-3 \left(C/\beta ^2\right) \sqrt{\phi } \right\}.
\end{split}
\end{equation}

Equation \eqref{xi}, \eqref{W} and \eqref{sigma} give the complete parametric solution of the field equations describing the geometry of a cosmic string in hybrid metric-Palatini gravity, with $\phi$ as a parameter. The three arbitrary integration constants appearing in the equations, $\xi _0$, $C$, and $W_0$, must be obtained from the initial or boundary conditions to be imposed on the cosmic string configuration. 

As for the mass of the string, in the dimensionless variable $\xi$ it can be obtained as
\begin{equation}
m(\xi)=2\pi \beta \int_0^{\xi _s}{\sigma (\xi)W(\xi)d\xi},
\end{equation}
where $\xi _s=R_s/\beta$.

In order to study specific cosmic string solutions in the hybrid metric-Palatini framework, we need now to specify the form of the potential $V(\phi)$. 

\section{Solutions to the dynamical equations with specific potentials}

In this section we will investigate the application of the set of parametric equations deduced on the previous section to different potential configurations, for a more complete set of possible potential configurations, we refer the reader to the original article Ref. \citenum{Harko:2020oxq}.

\subsection{Power law potential}

The gravitational field equations that describe a cosmic string in hybrid metric-Palatini gravity have an analytical solution, with a scalar field potential of the type $V(\phi)=V_0\phi ^{3/4}$. Rescaling the radial coordinate $r=\beta\xi$ and imposing the condition $\beta ^2V_0=1$, from Eq.~(\ref{xi}) we obtain the expression for the scalar field as a function of $\xi$, given by
\begin{equation}
\phi (\xi )=\frac{\left( \xi ^{2}\phi _{0}\,^{3/4}\pm 2\xi
\phi_{0}^{\prime}\pm 8\phi _{0}\right) ^{4}}{4096\phi
_{0}\,^{3}},
\end{equation}
where we made use of initial conditions $\phi (0)=\phi _0$ and $\phi '(0)=\phi _0^{\prime }$, respectively. As for metric tensor component $W$ we obtain 
\begin{equation}
W(\xi)=\frac{W_0}{\left(\xi ^2 \phi _0\,^{3/4}\pm 2 \xi  \phi_0^{\prime}\pm 8 \phi
   _0\right)^3 \sqrt{2 \xi \pm 2 \phi _0^{\prime}/\phi _0\,^{3/4}}}.
\end{equation}
where $W_0$ is an arbitrary constant of integration. On the string axis( $\xi =0$), we obtain $W^2(0)=\pm W_0\,^2/524288\, \phi _0\,^{21/4} \phi _0^{\prime}$. As the metric tensor component $W^2$ must be positive for all $\xi \geq 0$, it follows that the physical solution is the one with the positive sign. Hence in $V(\phi)=V_0\phi ^{3/4}$ potential, the solutions of the field equations describing a cosmic string in hybrid metric-Palatini gravity are
\begin{eqnarray}
\phi (\xi)&=&\frac{\left( \xi ^{2}\phi _{0}\,^{3/4}+ 2\xi \text{$%
\phi $}_{0}^{\prime }+ 8\phi _{0}\right) ^{4}}{4096\phi
_{0}\,^{3}}, \nonumber	\\
W^2(\xi)&=&\frac{W_0\,^2}{\left(\xi ^2 \phi _0\,^{3/4}+2 \xi  \phi_0^{\prime}+ 8 \phi
   _0\right)^6 \left(2 \xi + 2 \phi _0^{\prime}/\phi _0\,^{3/4}\right)},
\end{eqnarray}
respectively, with $W_0\,^2=524288\,W^2(0) \phi _0\,^{21/4} \phi _0^{\prime}$, a condition that implies $\phi _0>0$ and  $\phi _0^{\prime}>0$. For the string tension as a function of $\phi$ and $\xi$ we obtain the expressions
\begin{equation}
\kappa^2\sigma (\phi)=V_0\frac{-6 C-5(\phi -3) \sqrt[4]{\phi }}{8 \sqrt{\phi }},
\end{equation}
and

\begin{eqnarray}
\kappa^2\sigma (\xi)=&\frac{V_0 \phi _0^{3/4}}{\left(\xi ^2 \phi _0^{3/4}+2 \xi  \phi_0'+8 \phi _0\right)^2} 
 \Big\{-48C \phi _0^{3/4}
\quad -5 \left(\xi ^2\phi _0^{3/4}+2 \xi  \phi_0'+8 \phi _0\right)
 \times\nonumber \\& \left[\frac{(\xi ^2 \phi _0^{3/4}+2 \xi \phi _0'+8\phi_0)^4}{4096 \phi _0^3}-3\right]\Bigg\},
\end{eqnarray}
respectively.

In this particular potential configuration, the scalar field is a monotonically increasing function of the radial distance from the string axis and tends to infinity for $\xi \rightarrow \infty$. The metric tensor component, on the other hand, decreases monotonically from a finite value on the string axis and tends to zero at infinity. For $\xi =0$, the string tension takes the finite value 
\begin{equation}
\sigma (0)=V_0\left[-48 C \phi _0^{3/4}-40 \left(\phi _0-3\right) \phi _0\right]/64\phi _0^{5/4}
\end{equation}
while $\lim _{\xi \rightarrow \infty}\sigma (\xi)=-\infty$, indicating that $\sigma $ is a monotonically decreasing function of the radial coordinate. In the first order of approximation we obtain for the mass of the string of radius $\xi _s$ the expression

\begin{eqnarray}
m&=&\frac{\pi  \beta  W_0  \xi _s}{8192 \phi_0^{33/8}\phi _0^{\prime3/2}}\Big\{6 C \xi _s \phi_0^{7/4}+15 \xi _s \phi _0^{\;\prime \;2} \left(C-2 \sqrt[4]{\phi_0}\right)\nonumber \\ &-&4 \phi_0 \phi_0' \left[6 C+5 \left(\phi_0-3\right)
   \sqrt[4]{\phi_0}\right]+5 \xi _s \left(\phi_0-3\right) \phi_0^2\Big\}
   {}.
\end{eqnarray}
So, at first order, the mass monotonically increases with the string radius. 

\subsection{The Higgs-type potential}

Another type of scalar field potential studied is the Higgs type potential, given by
\begin{equation}
V(\phi) =\pm \frac{\bar{\mu}^2}{2}\phi ^2+\frac{\nu }{4}\phi ^4,
\end{equation}
where $\bar{\mu}^2$  and $\nu$ are constants. We will investigate only the case with $\bar{\mu}^2<0$, by adopting the minus sign in the above potential. The standard elementary particle physics approach assumes that the constant $\bar{\mu}^2$ is related to the mass of the scalar field particle as $m_{\phi}^2= 2\xi v^2 = 2\bar{\mu}^2$, where $v^2 = \bar{\mu}^2/\xi $ gives the minimum value of the potential. The Higgs self-coupling constant $\nu $ can be obtained from the experimental determination of the mass of the Higgs boson, and its numerical value is of the order of $\nu  \approx 1/8$ \cite{ATLAS:2015yey}. By rescaling the radial coordinate and the scalar field according to
\begin{equation}
r=\sqrt{2}\bar{\mu} \xi, \qquad  \phi=\frac{\Phi}{\left(\nu\bar{\mu}\right)^{1/3}},
\end{equation}

then Eq.~(\ref{xi}) provides the behaviour of the scalar field in the following form
\begin{equation}
\frac{d^2\Phi}{d\xi^2}-\frac{3}{4\phi}\left(\frac{d\Phi}{d\xi}\right)^2-\Phi ^2+\Phi^4=0.
\end{equation}

The general solution for this equation is given in a closed form  by
\begin{equation}
\xi +C_0=\int \frac{1}{\Phi ^{3/4} \sqrt{C+\frac{2}{21} \left(7-3 \Phi ^2\right) \Phi
   ^{3/2}}} \, d\Phi.
\end{equation}

However, this solution cannot be expressed in an analytical form in terms of known functions. In the first order approximation, we obtain
\begin{equation}
\xi +C_0\approx \frac{4 \sqrt[4]{\Phi }}{\sqrt{C}}-\frac{4 \Phi ^{7/4}}{21 C^{3/2}}+O\left(\Phi ^{9/4}\right),
\end{equation}

but this representation is not particularly useful from the point of view of concrete calculations. 

\begin{figure}[tbp]
 \centering
 \includegraphics[scale=0.65]{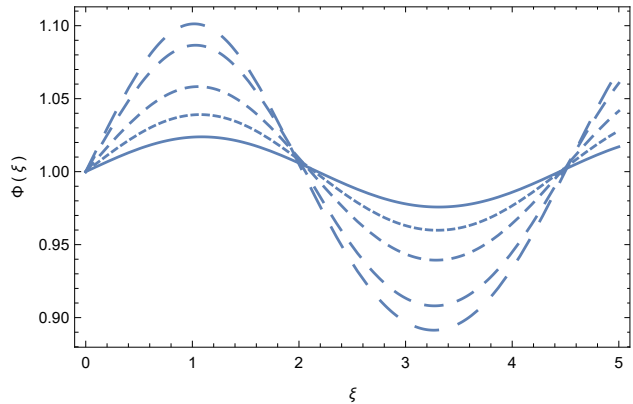}
 \caption[Variation of the scalar field for the cosmic string configuration in the presence of a Higgs-type potential]{Variation of the scalar field for the cosmic string configuration in the presence of a Higgs-type potential $V(\Phi)=-\Phi^2+\Phi^4$ for $\Phi (0)=\Phi_0=1$, and for different values of $\Phi_0'$: $\Phi_0'=0.034$ (solid curve), $\Phi_0'=0.056$ (dotted curve), $\Phi_0'=0.084$ (short dashed curve), $\Phi_0'=0.126$ (dashed curve), and $\Phi_0'=0.148$ (long dashed curve), respectively.}
 \label{fig7}
\end{figure}

The scalar field with Higgs potential supporting a string configuration in hybrid metric-Palatini gravity is represented in Fig.~\ref{fig7}.
Note that the scalar field for the Higgs-type potential shows a basically periodic structure, changing between successive maxima and minima, a very striking difference from other types of potential. There are singularities in the field, the profile is strongly dependent on the initial conditions at the string axis, and the field extends to infinity.

\begin{figure}[tbp]
 \centering
 \includegraphics[scale=0.65]{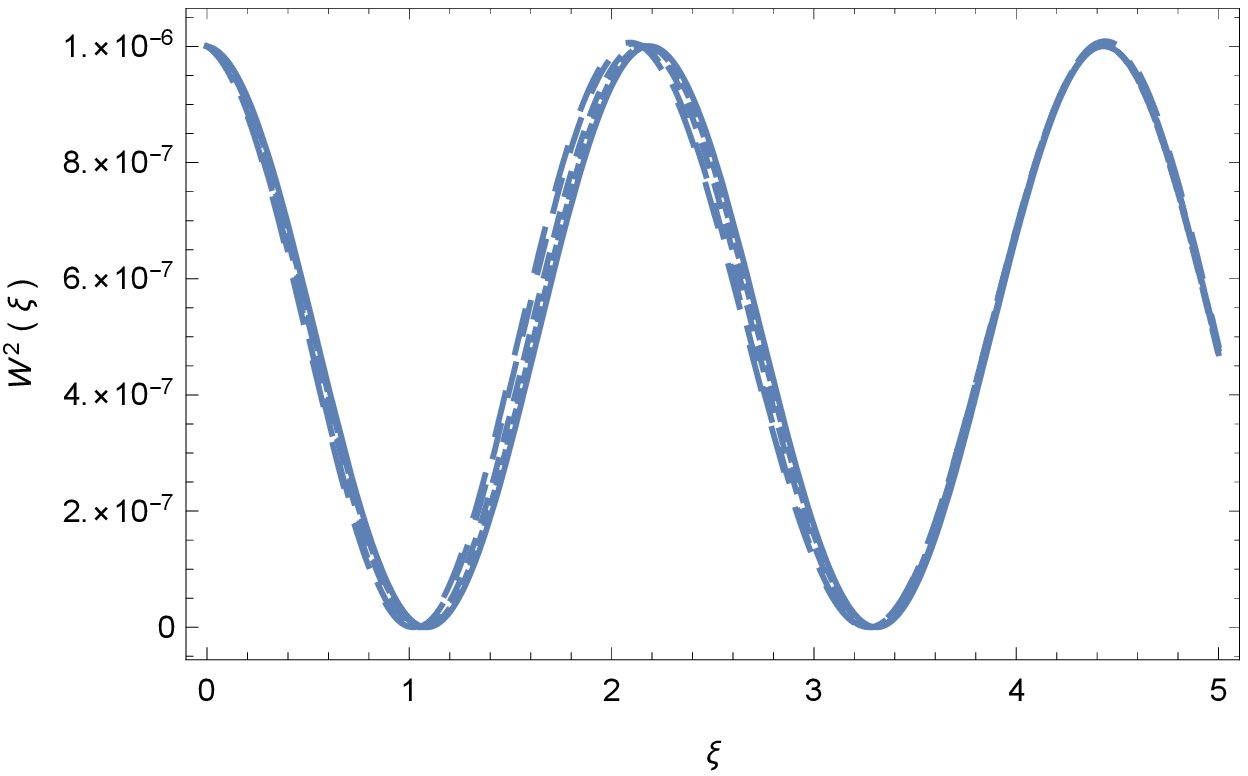}
 \caption[Variation of the metric tensor component $W^2(\xi)$ of the cosmic string configuration in the presence of a Higgs type potential]{Variation of the metric tensor component $W^2(\xi)$ of the cosmic string configuration in the presence of a Higgs type potential $V(\Phi)=-\Phi^2+\Phi^4$  for $\Phi \left(0\right)=1$, $W\left(0\right)=10^{-3}$, and for different values of $\Phi_0'$: $\Phi_0'=0.034$ (solid curve), $\Phi_0'=0.056$ (dotted curve), $\Phi_0'=0.084$ (short dashed curve), $\Phi_0'=0.126$ (dashed curve), and $\Phi_0'=0.148$ (long dashed curve), respectively.}
 \label{fig8}
\end{figure}

The variation of the metric tensor component $W^2(\xi)$ is represented in Fig.~\ref{fig8}.
The same oscillatory pattern found for the scalar field can also be observed for the metric tensor component $W^2$. However, there is a difference in the phase of these to quantities. When the field reaches its maximum at $\xi \approx 1$, the metric tends to zero, $W^2(1)\approx 0$. Then, while the scalar field decreases,  the metric tensor increases, reaching its maximum at the minimum of the field, corresponding to $\xi \approx 2$. This pattern is repeated up to infinity.

\begin{figure}[tbp]
 \centering
 \includegraphics[scale=0.65]{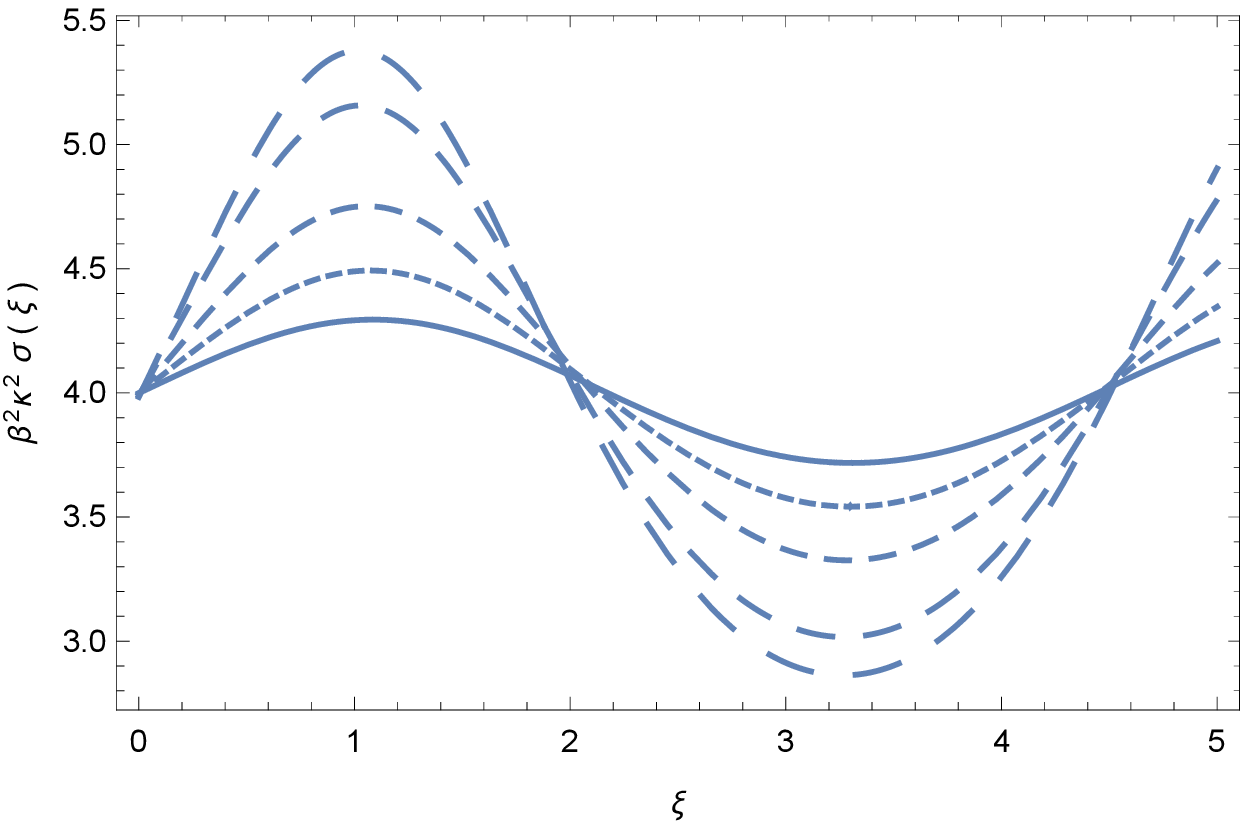}
 \caption[Variation of the string tension $\kappa ^2\sigma (\xi)$ of the cosmic string configuration in the presence of a Higgs type potential]{Variation of the string tension $\kappa ^2\sigma (\xi)$ of the cosmic string configuration in the presence of a Higgs type potential $V(\Phi)=-\Phi^2+\Phi^4$  for $\Phi \left(0\right)=1$, $W\left(0\right)=10^{-3}$, and for different values of $\Phi_0'$: $\Phi_0'=0.034$ (solid curve), $\Phi_0'=0.056$ (dotted curve), $\Phi_0'=0.084$ (short dashed curve), $\Phi_0'=0.126$ (dashed curve), and $\Phi_0'=0.148$ (long dashed curve), respectively.}
 \label{fig9}
\end{figure}

The variation of the string tension with the radial coordinate is depicted in Fig.~\ref{fig9}. We can see the same oscillatory behaviour, and  $\sigma$ varies in phase with the scalar field. This oscillatory pattern is a general property of all physical and geometrical parameters when fields with Higgs-type potential are considered.

\section{Conclusions}

In this work we studied the existence and physical properties of cosmic strings in the context of the hybrid metric-Palatini gravity. The theory is an extension to General Relativity, combining both metric and Palatini formalism. A main success of the theory is the possibility to generate long-range forces that pass the classical local tests of gravity at the Solar System level, thus avoiding some problematic features of the standard $f(R)$ theories. Another interesting advantage of the theory is that it admits an equivalent scalar-tensor representation, simplifying greatly the dynamical equations. 
The type of strings studied in this work are local gauge strings, using an approximation to the Vilenkin-prescribed energy-momentum tensor and different potential configurations.
For the case of the potential of the form $V=V_0\phi^{3/4}$. The requirement of the positivity of the metric tensor for $r=0$ imposes the condition that the metric tensor and the string tension are decreasing functions of $r$, since both $\phi_0$ and $\phi'_0$ are required to be positive. In this case, the scalar field becomes singular at infinity. However, it is possible to construct finite string configurations: defining a string radius by introducing an effective cuttoff length, $\xi_{co}$ for both the metric and the scalar field. This radius would allow us to get finite values for the scalar field, the metric tensor components and the string tension. For this type of potential the choice of $\xi_{co}$ should be made based on empirical considerations, such as consistency with observational data.
The case of the Higgs-type potential is quite different, since the string tension does not vanish for any value of the radial coordinate but it does reaches its minimum value at $\xi\approx3.5$, where the scalar field is also at its minimum and $W^2(\xi)$ is singular and tends to zero. Another possible choice for the string radius could be $\xi\approx 1$, the first zero of the metric tensor component; in this case both the string tension and the scalar field are at their maxima. Or we could, as in the exponential potential, introduce a cutoff radius to be determined by confrontation with observations, matching the solution with the cosmological background or another GR string solution, thus achieving a well behaved structure throughout the radius of the string.

\section{Citations}

\section{Acknowledgments}
F. S. N. L. acknowledges support from the Fundação
para a Ciência e a Tecnologia Scientific Employment
Stimulus contract with reference CEECIND/04057/2017
and the research Grants No. UID/FIS/04434/2019 and
No. PTDC/FIS-OUT/29048/2017.

\end{document}